\documentclass[fleqn,usenatbib,referee]{mnras}

\usepackage[T1]{fontenc}

\DeclareRobustCommand{\VAN}[3]{#2}
\let\VANthebibliography\thebibliography
\def\thebibliography{\DeclareRobustCommand{\VAN}[3]{##3}\VANthebibliography}

\usepackage{graphicx}

\usepackage{amsmath}    % Advanced maths commands
\usepackage{amssymb}    % Extra maths symbols

\title[Refining the prediction for OJ~287 next flare]{Refining the prediction for OJ~287 next impact flare arrival epoch}

\author[M. Valtonen et al.]{Mauri J. Valtonen,$^{1,2}$\thanks{E-mail: mvaltonen2001@yahoo.com (MJV)} 
Staszek Zola,$^3$
A. Gopakumar,$^4$
Callum McCall,$^5$
Helen Jermak,$^5$ 
\newauthor{Lankeswar Dey,$^4$ S. Komossa,$^6$ Tapio Pursimo,$^7$ Emil Knudstrup,$^7$ Dirk Grupe,$^8$} 
\newauthor{Jose L. Gomez,$^9$ Rene Hudec,$^{10,11}$ Martin Jel\'{\i}nek,$^{11}$  Jan \v{S}trobl,$^{11}$ Andrei V. Berdyugin,$^2$}
\newauthor{Stefano Ciprini,$^{12,13}$ Daniel E. Reichart,$^{14}$ Vladimir V. Kouprianov,$^{14}$ Katsura Matsumoto,$^{15}$}
\newauthor{Marek Drozdz,$^{16}$ Markus Mugrauer,$^{17}$ Alberto Sadun,$^{18}$
Michal Zejmo,$^{19}$ Aimo Sillanp\"a\"a,$^2$}
\newauthor{Harry J. Lehto$^2$ and Kari Nilsson$^1$}
\\
$^1$ FINCA, University of Turku, Turku, Finland\\
$^2$ Tuorla Observatory, Department of Physics and Astronomy, University of Turku, Turku, Finland\\
$^3$ Astronomical Observatory, Jagiellonian University, ul. Orla 171, 30-244 Krakow, Poland\\
$^4$ Department of Astronomy and Astrophysics, Tata Institute of Fundamental Research, Mumbai, India\\
$^5$ Astrophysics Research Institute, Liverpool John Moores University, Liverpool Science Park IC2, 146 Brownlow Hill, UK\\
$^6$ Max-Planck-Institut f\"ur Radioastronomie, Auf dem H\"ugel 69, 53121 Bonn, Germany\\
$^7$ Nordic Optical Telescope, Apartado 474, E-38700 Santa Cruz de La Palma, Spain\\
$^8$ Department of Physics, Geology, and Engineering Technology, Northern Kentucky University, 1 Nunn Dr, Highland Heights, KY 41076, USA\\
$^9$ Instituto de Astrofisica de Andalucia - CSIC, Glorieta de la Astronomia s/n, 18008 Granada, Spain\\
$^{10}$ Czech Technical University, Faculty of Electrical Engineering, Prague, Czech Republic\\
$^{11}$ Astronomical Institute (ASU CAS), Ond\v{r}ejov, Czech Republic\\
$^{12}$ Instituto Nazionale di Fisica Nucleare (INFN) Sezione di Roma Tor Vergata, Via della Ricerca Scientifica 1, 00133, Roma, Italy\\
$^{13}$ ASI Space Science Data Center (SSDC), Via del Politecnico, 00133, Roma, Italy\\
$^{14}$ University of North Carolina at Chapel Hill, Chapel Hill, North Carolina, NC 27599, USA\\
$^{15}$ Astronomical Institute, Osaka Kyoiku University, 4-698 Asahigaoka, Kashiwara, Osaka, 582-8582, Japan\\
$^{16}$ Mt. Suhora Observatory, Pedagogical University, ul. Podchorazych 2, 30-084 Krakow, Poland\\
$^{17}$ Astrophysikalisches Institut und Universitäts-Sternwarte, Schillergässchen 2, D-07745 Jena, Germany\\
$^{18}$ Department of Physics, University of Colorado, Denver, CO 80217, USA\\
$^{19}$ Kepler Institute of Astronomy, University of Zielona Gora, Lubuska 2, 65-265 Zielona Gora, Poland
}

\date{Accepted ... Received ... in original form ...}

\pubyear{2022}

\begin{document}
\label{firstpage}
\pagerange{\pageref{firstpage}--\pageref{lastpage}}
\maketitle

%\author[0000-0003-3609-382X]{Staszek Zola}
%\author[0000-0003-4274-4369]{A. Gopakumar}
%\author[0000-0002-3375-3397]{Callum McCall}
%\author[0000-0002-1197-8501]{Helen Jermak}
%\author[0000-0002-2554-0674]{Lankeswar Dey}
%% Stefanie: prefer to go without orcid, and NEVER spell out my first name.
%\author[0000-0002-9214-4428]{S. Komossa}
%\author[0000-0002-9961-3661]{Dirk Grupe}
%\author[0000-0003-4190-7613]{Jose L. Gomez}
%\author[0000-0002-7273-7349]{Rene Hudec}
%\author[0000-0003-3922-7416]{Martin Jel\'{\i}nek}
%\author[0000-0002-4147-2878]{Jan \v{S}trobl}
%\author[0000-0002-9353-5164]{Andrei V. Berdyugin}
%\author[0000-0002-0712-2479]{Stefano Ciprini}
%\author[0000-0003-3642-5484]{Vladimir V. Kouprianov}
%\author[0000-0002-5277-568X]{Katsura Matsumoto}
%\author[0000-0001-5836-9503]{Michal Zejmo}
%\author[0000-0002-1445-8683]{Kari Nilsson}

% Abstract of the paper

\begin{abstract}
The bright blazar OJ~287 routinely parades high brightness bremsstrahlung flares which are explained as being a result of a secondary supermassive black hole (SMBH) impacting the accretion disk of a primary SMBH in a binary system. We begin by showing that these flares occur at times predicted by a simple analytical formula, based on the Kepler equation, which explains flares since 1888. The next impact flare, namely the flare number 26, is rather peculiar as it breaks the typical pattern of two impact flares per 12 year cycle.  This will be the third bremsstrahlung flare of the current cycle that follows the already observed 2015 and 2019 impact flares from OJ~287. Unfortunately, astrophysical considerations make it difficult to predict the exact arrival epoch of the flare number 26.  In the second part of the paper,  we describe our recent OJ~287 observations. They show that the pre-flare light curve of flare number 22, observed in 2005, exhibits similar activity as the pre-flare light curve in 2022, preceding the expected flare number 26 in our model. We argue that the pre-flare activity most likely arises in the primary jet whose activity is modulated by the transit of the secondary SMBH through the accretion disk of the primary. Observing the next impact flare of OJ~287 in October 2022 will substantiate the theory of disk impacts in binary black hole systems.
\end{abstract}


\section{Introduction} \label{sec:intro}

Supermassive black hole (SMBH) binary systems are expected in the standard cosmological scenario as most massive galaxies contain a SMBH at their center and binaries should form by the merger of these galaxies \citep{BBR80,val89,mik92,val96b,qui96,mil01,vol03,kz2016,bur18}. Electromagnetic observations suggest the existence of a couple of dozen SMBH binary candidates in active galactic nuclei \citep{charisi2016,Graham2015,Zhu2020,Bon2016,Liu2014, lai99,kau17,Iguchi2010}.

However, detailed theoretical investigations and observational campaigns make OJ~287, a BL Lacertae object at a redshift of 0.306 \citep{sit85,nil10}, a very special SMBH binary candidate \citep{val21}. Interestingly, the binary nature of OJ~287 central engine was recognized by one of us (Aimo Sillanp\"a\"a) already back in 1982, while constructing historical light curves for the quasars in the Tuorla - Mets\"ahovi variability survey which had begun two years earlier \citep{kid07}. This inference was based on the observational evidence for major flares around 1911, 1923, 1935, 1947, 1959 and 1971 in the historical light curve of OJ~287. From this sequence it was easily extrapolated that OJ~287 should display a major outburst in 1983. The blazar monitoring community was alerted, resulting in a successful observational campaign of OJ~287. Indeed, one of biggest flares ever observed in OJ~287 occured at the beginning of 1983 \citep{sil85,smi85}.

Following this success, further flares were predicted by \cite{sil88}, the next one in the autumn of 1994. It was indeed verified by the second campaign called OJ-94 \citep{sil96a}.

It was recognised soon after that these flares in OJ~287 were not exactly periodic, and that the systematics of the past flares are better understood if the flares are double-peaked and the two peaks are separated by $\sim 1-2$ years \citep{val96,LV96}. This led to the proposal of a new SMBH binary  central engine model for OJ~287 where the secondary SMBH orbits the more massive primary SMBH in a relativistic eccentric orbit with a redshifted orbital period of $\sim 12$ years. The orbital plane is inclined with respect to the accretion disk of the primary at a large angle which leads to the secondary SMBH impacts with the accretion disk of the primary twice every orbit. These impacts lead to double-peaked flares in OJ~287. The next campaign carried out by the OJ-94 group verified the flare on October 1995, the second one of the pair. Interestingly, it came within the narrow two-week time window of the prediction \citep{val96,sil96b}. The property of paired flares could be called symmetry, in terminology often used in physics. 

 Subsequently, a number of investigations were pursued to improve astrophysical, observational and theoretical aspects of the SMBH binary central engine description for OJ~287  \citep{Ram07,val10,Hudec2013,Pursimo2000,val06,val06a,val08,val11b,val16,Laine20,dey19}. These efforts allowed us to obtain the following values for OJ~287's SMBH binary system:
primary mass $m_1 = 18.35 \times 10^9 M_{\odot}$, secondary mass $m_2 = 150 \times 10^6 M_{\odot}$, primary Kerr parameter $\chi_1 = 0.38$, orbital eccentricity $e = 0.657$, and orbital period (redshifted) $P = 12.06$ years \citep{Dey18}. These are among the nine parameters of a unique mathematical solution which may be obtained if the timing of ten optical outbursts are known. The solution exits only if each of the ten flares come within a narrow time window whose width is specified in  \cite{Dey18}.

Further, it turns out that the up-to-date SMBH binary orbit solution is consistent with additional seven flare epochs \citep{dey19}. Thus the solution is strongly over-determined. The resulting impact flare epoch sequence, extracted from \cite{Dey18}, reads  
1886.62 (1), 1896.67 (1), 1898.61 (3), 1906.20 (4), 1910.59 (5), 1912.98 (6), 1922.53 (7), 1923.73 (8), 1934.34 (9), 1935.40 (10), 1945.82 (11), 1947.28 (12), 1957.08 (13), 1959.21 (14), 1964.23 (15), 1971.13 (16), 1972.93 (17), 1982.96 (18), 1984.12 (19), 1994.59 (20), 1995.84 (21), 2005.74 (22), 2007.69 (23), 2015.87 (24), 2019.57 (25), 2022.55 (26) 
where we use brackets to denote the sequence number. Among the flares in the list, eight flares are only suspected as there are not enough historical observations for our timing purposes even though there are indications that impact flares might have happened from after-flare activities. It is also important that there are no known flares in the historical light curve that would invalidate the above sequence.

One of the details that goes into the calculation of the flare epochs is the fact that the accretion disk does not stay exactly at its mean plane, but bends slightly (of the order of 1 degree) on either side of it due to the tidal influence of the secondary black hole. It is taken care by a single-valued function of the distance of the impact point from the primary SMBH \citep{val07}.  It affects the time of the appearance of the flare by a small amount.

However, an exception occurs concerning the flares $\#\ 22$ and $26$. Numerical simulations show that the disk bending is quite different during the SMBH impact epochs associated with these two cases and actually in opposite directions even though the distances of their impact sites from the center are roughly the same \citep{val07}. Specifically, these  simulations place the disk $265\pm80$ AU above the mean plane during 2005, while in 2022 the disk plane is $235\pm80$ AU below the mean disk plane (\cite{val07}, Table 1). Additionally, the disk bending in 2022 is sensitive to the precession rate of the orbit as noted in \cite{val07}. These considerations imply that the 2022 flare should happen at a later epoch than estimated in \cite{Dey18} who used the disk level of 2005 to calculate the 2022 flare time. Quantitatively, extrapolating between the two precession models in \cite{val07} to the latest precession value of \cite{Dey18}, we expect that the time difference between the disk crossings during 2005 and  2022 should be $17.04\pm0.03$ yr and the impact flare epoch should happen in October 2022.
This may be contrasted with the epochs of crossings of the average midplane of the accretion disk which occur at 2005.242 and at 2022.055, respectively, and which lead to a time difference of 16.813 years between the impacts \citep{Dey18}.

Let us note that these disk bending uncertainties do not influence the secondary SMBH orbit solution in any way since uncertainties do not appear in the ten flares used for the unique mathematical solution, such as the Einstein centenary flare \citep{val10,val16}, or the Eddington flare \citep{Dey18,Laine20}.  In the latter case, no disk level corrections were necessary, and in any case corrections would have been identical, and cancelled out in calculating the time difference between the two very similar periastron flares in 2007 and 2019.

Typically the first optical data are obtained around the fraction .65 of any year, after a short summer break, when the elongation of OJ~287 from the sun is too small for ground based optical observations. With the 16.813 yr time difference between 2005 and 2022 flares, OJ~287 would be already undergoing a flare when the observations resume after the summer break in 2022 \citep{val21}. On the other hand, the 17.04 yr time difference puts OJ~287 in a low state at this time. Thus already the first observation of the OJ~287 in the latter part of 2022 is able to give us a preliminary answer with regard to the disk level in 2022. Note that the flares do not have a counterpart in radio or X-ray region where it is possible to get data at smaller solar elongations.
  
The two nearly identical SMBH crossing configurations during 2005 and 2022 should generate very similar light curves from the source. It has been previously demonstrated that the disk crossing influences the optical flux in a predictable way \citep{sun96,sun97,val09,val11a,val17}, and it is reasonable to expect that two nearly identical crossings produce similar light curves. This assumption will be studied below. 

The paper is organised as follows. We begin by providing a simple analytical formula to obtain the first order epochs of these bremsstrahlung flares. We call this Keplerian sequence. Then we mention briefly how the second order model produces more accurate predictions of these impact flare arrival epochs. Thereafter, we describe the recent observational campaigns to narrow down the epoch of the recent secondary SMBH impact. Their implications for the arrival of the next impact flare are provided in Section 4.

\section{Predicting Impact flare arrival Epochs}
 We begin by providing a mathematical prescription for  providing a sequence of epochs that is fairly close to the one we displayed earlier. This prescription arises essentially from the celestial mechanics and GW phasing considerations and is bereft of astrophysical inputs \citep{valkar06,TG07}.
Thereafter, we clarify why astrophysical considerations are crucial for  accurately predicting the epochs of   impact flare arrivals 

\subsection{The first order ephemeris of flare times: A Keplerian Sequence}

 We term the mathematical prescription that provides
 a first description of the arrival epochs of impact flares as a Keplerian sequence.
 This is due to the use of the classical Kepler equation that connects the eccentric anomaly $u$ to the mean anomaly $l$ \citep{valkar06}
 \begin{equation}
     l = u -e \, \sin u  \,,
 \end{equation}
 where $l = 2 \, \pi/ T_{orb}$  and $T_{orb}$ being the orbital period and $e$ its orbital eccentricity, and $u$ and the phase angle $\phi$ are connected by standard formulae.

The Keplerian sequence which is useful in understanding OJ~287's impact flares is characterised by an orbital period 12.13 years, eccentricity e = 0.65, forward precession $\Delta \phi = 38^{\circ}$ degrees per period and the initial angle from the pericenter to the fixed line $+1^{\circ}$ at the year 1910.50, one of the moments of pericenter.  Every time the particle moves over the fixed line, the phase angle of the fixed line jumps down by $\Delta \phi$, thus mimicking forward precession of the major axis of our elliptical orbit.  The ephemeris of conjunctions is then easily calculated by using formulae in \cite{valkar06}.
We start from the pericenter times 
\begin{equation}
T_{p}(n) = 1874.11 + 12.13 n, n=1, 2, 3...
\end{equation}

where $n$ is the orbit number, and from Kepler’s Equation written as a function of the phase angle $\phi_i(n)$:

%SZ  equation below should be split into two lines
\begin{equation}
\begin{split}
T(\phi_i(n)) =  12.13/2\pi  (2 arctan(0.46 tan(\phi_i(n)/2)) - 0.598 tan(\phi_i(n)/2)  \\ /(1 + 0.2116 tan^2(\phi_i(n)/2)))
\end{split}
\label{Eq_KE_v}
\end{equation}

 where $\phi_i(n)$ is the phase angle at the crossing of the line of nodes \citep{valkar06}. Its values $\phi_i(n), i = -1, 0,+1$, come from the set of first flare phase angles $\phi_1(n)$, $n$ = 2,…,12, second flare phase angles $\phi_0(n)$,  $n$ = 1,…,12, and occasional third flare phase angles $\phi_{-1}(n)$, $n$ = 3,7,12:

\begin{equation}
\phi_1(n)  = (257 - 38 (n-1)) ^{\circ}, n = 2,3,8,...,12
\end{equation}
\begin{equation}
\phi _1(n) = (77 - 38 (n-1)) ^{\circ}, n = 4,...,7
\end{equation}
\begin{equation}
\phi _0(n) = (77 - 38 n) ^{\circ}, n = 1,2,8,...,12
\end{equation}
\begin{equation}
\phi_0(n)  = (257 - 38 n) ^{\circ}, n = 3,...,7
\end{equation}
\begin{equation}
\phi_{-1}(3) = 1^{\circ}
\end{equation}
\begin{equation}
\phi_{-1}(7) =  171^{\circ}
\end{equation}
\begin{equation}
\phi_{-1}(12) = 161^{\circ}
\end{equation}

from which the line-crossing times are

\begin{equation}
T_1(n) = T_{p}(n) - T(\phi_1(n))
\end{equation}
\begin{equation}
T_0(n) = T_{p}(n) + T(\phi_0(n))
\end{equation}
\begin{equation}
T_{-1}(n) = T_{p}(n) + T(\phi_{-1}(n))
\end{equation}

This produces a list of times with a sequence number $k = 2n-i-1$ for $k = 1,...,4, k = 2n-i$ for $k = 6,...,15$ and $k = 2n-i+1$ for $k = 16,...,26$. The sequence number $k = 5$ arises when $n = 3$ and $i = -1$. Thus the list starts $T_2(1), T_1(2), T_2(2), T_1(3), T_3(3), T_2(3), T_1(4), T_2(4), T_1(5),...$ or:
1886.49 (1), 1897.05 (2), 1898.38 (3), 1904.56 (4), 1910.51 (5), 1912.95 (6), 1922.42 (7), 1923.61 (8), 1934.24 (9), 1935.20 (10), 1945.72 (11), 1947.05 (12), 1958.12 (13), 1958.97 (14), 1964.01 (15), 1971.10 (16), 1973.04 (17), 1983.00 (18), 1984.07 (19), 1994.77 (20), 1995.77 (21), 2006.06 (22), 2007.64 (23), 2015.76 (24), 2019.57 (25), 2023.59 (26),....

A close inspection reveals that 
the sequence of epochs obtained in this way is rather close to the list of epochs that arise  from the
SMBH binary central engine description \citep{Dey18}. 
Specifically, the triplet epochs, namely  2015.76, 2019.57 and 2023.59 in our Keplerian sequence 
closely follow the epochs  
2015.87, 2019.57 and 2022.55 that arise from 
the full mathematical solution.

However, we need a higher order solution of the orbit as well as estimates of various astrophysically relevant delays. Using them, we get the second order ephemeris for OJ~287.

Few comments are in order before we clarify the astrophysical difficulties in estimating the arrival epoch of the flare \#26. 
Note that the requirement that phase angle $\phi_i$ should take 3 sets of values is reminiscent of the way frequencies are distributed in the GW spectrum of non-spinning BH binaries in relativistic/precessing eccentric orbits \citep{TG07}.
Recall that GWs are emitted at integer multiples of the orbital frequency for BH binaries in Newtonian eccentric orbits \citep{PM63}.
This essentially arises from the Fourier-Bessel series expansion of the Newtonian eccentric orbit in terms of the mean anomaly $l$ \citep{valkar06}.
However, each GW spectral lines splits in a triplet 
when the effects of periastron advance is included \citep{TG07}.
In other words, the frequency $ f_n \rightarrow ( f_n, f_{n \pm \delta f})$ where $\delta f = 4\, \pi\,k\,f_{orb}$
where $k$ the rate of periastron advance.
This structure essentially arises from the fact that there are two timescales that are associated with the orbital period and the periastron advance.
A similar structure in our Kepler sequence prescription naturally arises as we provide fixed angular jumps $\Delta \phi$ to the phase angle at certain fixed lines that mimic, as noted earlier, the effects of periastron advance.  We now briefly explain astrophysical delays that are present in our SMBH binary central engine description for OJ~287.

\subsection{The second order ephemeris of OJ~287 flares}

It is hard to imagine that there could be any other reason for creating flares in OJ~287 in a Keplerian sequence other than having components which follow a nearly Keplerian motion, such as what happens in a binary black hole system. Alternative suggestions have not led to correct predictions of future flares, nor to a full explanation of the previously known flares \citep{vil98,rie04,dey19}.

However, in a binary black hole system we need the Post-Newtonian corrections and to include the influence of the primary spin \citep{KG2005}. This will generate a new sequence which is not quite the same as the Keplerian sequence, but not very different from it either.

Moreover, the fixed line has to be represented by an astrophysical entity, in this case the line of nodes between the accretion disk plane and the orbital plane. Additionally, we must consider the process that generates the flares at the crossing of the line of nodes \citep{iva98}. Necessarily a time delay $t_{del}$ arises between the line crossing and the flare \citep{LV96}. Since the distance from the center varies from one crossing to the other, the time delays also vary. The delay may be calculated in a standard model of the accretion disk \citep{val19}. The standard model is described by two parameters; thus two more flare times are required for the unique solution of the sequence. In all we need 10 flare times, as mentioned above, to generate the second order ephemeris of flares. This is the sequence given in the introduction, where the member $\#26$ is 2022.55, rather than 2023.59 of the Keplerian sequence.

For the next order of accuracy we would need the exact level of the accretion disk at the 2022 disk impact. As mentioned in the introduction, there is a way to extrapolate the information from \cite{val07}. However, a new disk simulation should be done with the correct value of orbit precession. This is a laborious project and has to be left for further work. In the meantime, we turn to observations which may give us the same information directly, without knowledge of the disk level.

\section{Observations}

\subsection{Optical data}
\label{sec:optical}

\begin{figure*}
\includegraphics[angle=270,scale=0.62]{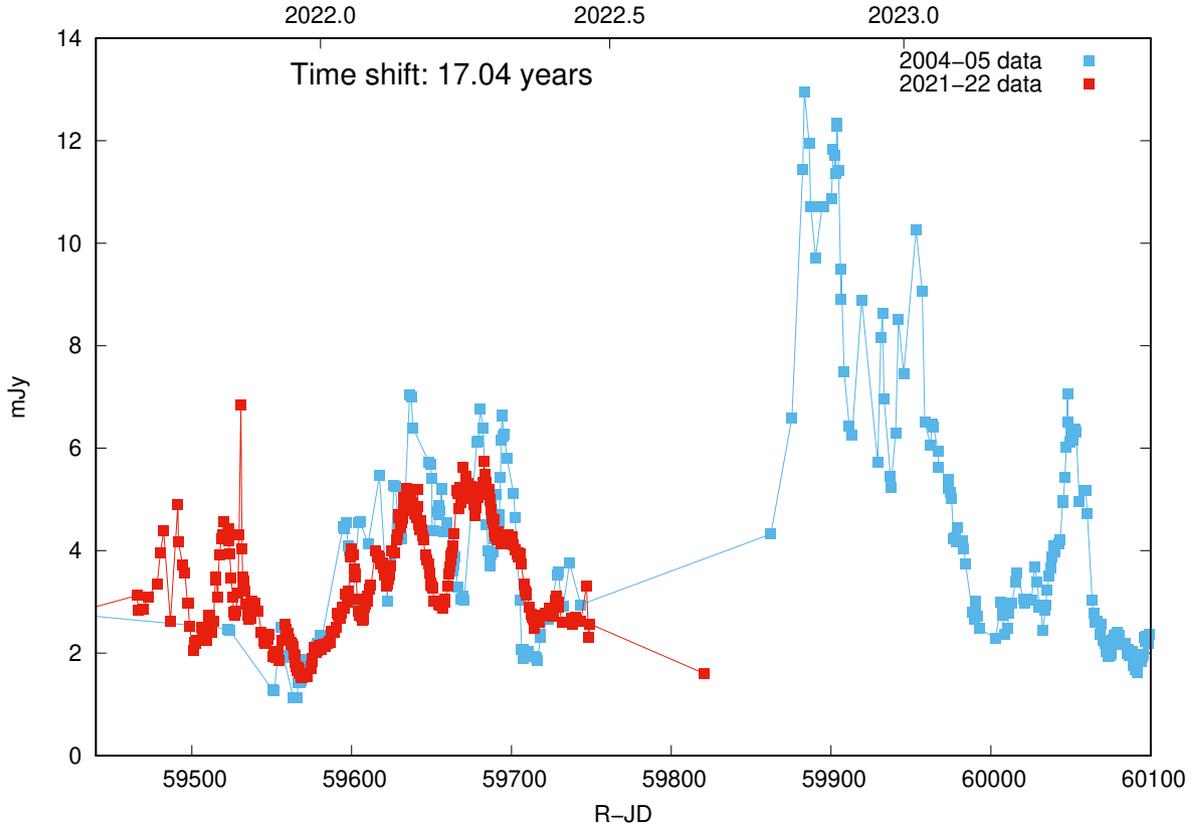}
\centering
\caption{The R-band light curve of OJ287 in the 2021/22 observing season superimposed on the 2004/05 one with 17.04 years shift applied. The most recent observations were taken on Aug 28, 2022, on JD = 2459820.309 when R mag = 15.49 $\pm$ 0.09.}
\label{lc05_22}
\end{figure*}

Optical data presented in this work consist of historical set gathered in the wide band R filter \citep{val06a,wu06,ciprini2007} and a recent R filter dataset taken within the Krakow Quasar Monitoring Program. The latter consists predominantly of observations obtained with the Skynet Telescope Robotic Network \citep{zola21}, appended with points from other telescopes at Osaka, Krakow, Jena, Mt. Suhora and  Ond\v{r}ejov Observatories \citep{mugrauer2010,mugrauer2016} . The location of telescopes on four continents and their redundancy allowed to achieve daily sampling, often we were able to collect data twice a day, if needed. Altogether 45217 single points have been collected since the start of the 2015/16 observing season. Binning them with half a day results in  2315 mean points. Observations discussed here cover September, 2021 to June, 2022 period and contain 315 mean points, shown by red squares in  Figure \ref{lc05_22}.

OJ~287 was well covered by optical photometry during the 2004/05 season. The points in the 2004/5 light curve are 0.01 yr averages from over 4000 single photometric observations. After a deep minimum in December 2004 there was a rather steady rise in brightness up to February 2005 maximum, followed by another maximum in April 2005. The 2021/22 light curve is surprisingly similar: after a minimum in December 2021 the brightness increased to a maximum in February 2022 with a second maximum in April. The 2021/22 light curve, superimposed on the 2004/05 light curve with the expected 17.04 yr time difference, is shown in Figure \ref{lc05_22}.

\subsection{Swift data}
%\label{sec:swift}

Nasa's Neil Gehrels Swift observatory was used to study OJ~287~ in the course of the project MOMO (Multiwavelength Observations and Modelling of OJ~287; \cite{Komossa2021b}). In this project, the two narrow-field telescopes aboard Swift are utilised: the UVOT and the XRT, which includes all six Swift optical and UV filters (17-600nm), and the X-rays (0.3-10keV).  The cadence ranges between typically 5 days (at inactive states) and 1 day (at outburst states or other states of particular interest). An analysis of timing and spectral properties of OJ~287 at all states of activity until January 2022 has been presented in a sequence of publications \citep{Komossa2020,Komossa2021a,Komossa2021b,Komossa2021c,Komossa2021d,Komossa2022a}. The data mentioned here cover the time interval 2021 October to 2022 March.

Another clue to the origin of optical activity of OJ~287 at this time is given by the continuum spectrum. For most times the spectrum from optical to infrared has a rather constant spectral index independent of level of activity \citep{kid18}.
In the optical--UV, dense monitoring of OJ~287 with Swift between 2015--2022 has shown that the optical and UV are always closely correlated but the flux ratio $F_{\rm V}/F_{\rm W2}$ can vary very significantly up to a factor of $\sim$2 \citep{Komossa2021d}, with a pattern that the spectrum is bluer when brighter during outbursts \citep{Komossa2021b,Komossa2022b}.
However, a major deviation from the relatively constant IR--optical shape happens during impact flares when the additional emission has a flat spectrum, causing the overall spectrum to flatten also \citep{val12,Laine20}. At the other end of brightness, during very deep fades the host galaxy contribution makes the IR--optical spectrum steeper \citep{val22}. The X-ray emission of OJ 287 is closely correlated with the optical--UV only during major outbursts (most recently in 2016/17 and 2020), but shows less or no correlation during more quiescent states \citep{Komossa2021d} and the fractional amplitude of variability, $F_{\rm var}$, is lower in X-rays than in the optical--UV \citep{Komossa2022b}.

\begin{figure*}
\includegraphics[width=\textwidth]{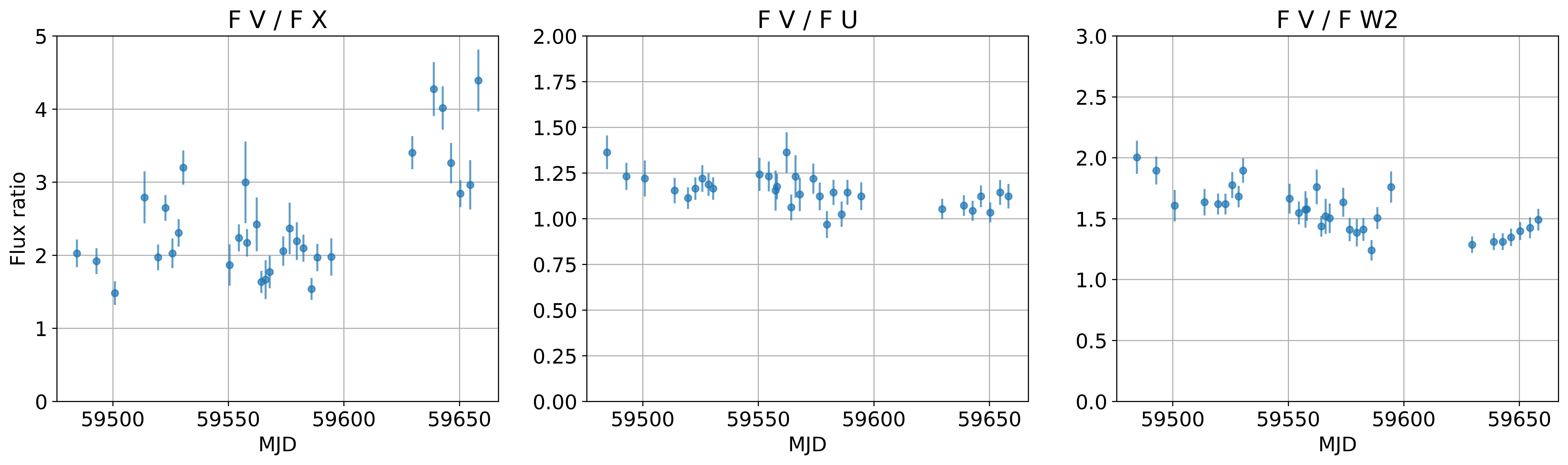}
\centering
\caption{Spectral ratios in the Swift data between the visual flux $F_V$, near-ultraviolet flux $F_U$, far ultraviolet flux $F_{W2}$ and X-ray flux $F_X$. The main optical flaring occurs to the right of JD 2459600 but since there is no corresponding X-ray flaring, the spectral ratio between the visual flux $F_V$ and the X-ray flux $F_X$ jumps to a higher value. The same pattern of variability has been observed at other epochs including in late 2020--early 2021. The other ratios remain rather constant.} 
\label{swift_spec}
\end{figure*}

Figure \ref{swift_spec} shows spectral ratios in selected bands in the optical, UV and X-rays during the epoch of interest between October 2021 and March 2022.

\section{Discussion and Conclusions}
%\label{Sec_Disc}

In the latest orbit model the corresponding  crossings of the mid-plane of the system happen at 2005.242 and at 2022.055, respectively, 16.813 yr apart from each other \citep{Dey18}. This is 0.23 yr less than the expected difference between the thermal flares mentioned in the Introduction, and which was also used in overlaying the data sets in Fig. \ref{lc05_22}. The orbit model has been well confirmed \citep{Laine20}, so the difference most likely has to do with the plane of the accretion disk, which has changed between 2005 and 2022. From our viewing direction, the disk plane was closer to us in 2005 than it is in 2022.

The first observations of the fall season of 2022 have just been carried out in Osaka (K.M.) while this paper was written. OJ287 was found in a low state, as shown in Figure \ref{lc05_22}. This excludes the possibility of the 16.813 yr time delay which would have put OJ287 in high state \citep{val21}.

In the model of \cite{Dey18} the impact on the upper surface of the accretion disk happens 460 AU above the mean disk (i.e. the disk is bent upwards by $\sim$1 degree; here "up" means toward the observer). 
The 17.04 yr time shift implies that the bending was downwards in 2022 by a similar amount, the impact point lying 350 AU below the mean disk position. This is easy to understand qualitatively, since during the previous half-an-orbit the secondary pulled the accretion disk from above in 2005 and from below in 2022. The disk simulations have shown that at least qualitatively this is correct \citep{val07}. 

The time of the primary flare is obtained from the disk impact time by adding the quantity $t_{del}$. Since we have no reason to expect that the standard value $t_{del}$ = 0.62 yr at this impact distance would need to be changed, the time of the primary flare becomes $t_{out} = 2022.55 + 0.23 = 2022.78$, that is October 10, 2022, $\pm$ 10 days. The value of $t_{del}$ = 0.62 yr cannot be very much wrong because it leads to the correct time for the 2005 flare. 

What is the nature of this pre-flare radiation? We have found that the radiation has polarisation properties which are no different from ordinary flares. The degree of polarisation rises with increasing brightness, as for example in the 2016/17 after-flares \citep{val17}. These are associated with the primary jet.

The position angle of the primary jet, as determined from jet model \cite{Dey21} as well as from VLBI observations \citep{Gom22}, agrees with the polarisation position angle observed in this work. The models give $PA = 123 - 128^{\circ}$ \citep{val21}. The values should be measured at a quiescent state, before or after a flare. We find that at these times $PA \sim 125\pm15^{\circ}$. These data will appear in more detail in a later publication (C.M., H.J., A.B.). For the secondary jet, the position angle should be $\sim193^{\circ}$ \citep{Dey21,val21} which is quite different.

The optical-UV and X-ray Swift data of OJ 287 taken in the course of the MOMO project
\citep{Komossa2020,Komossa2021a,Komossa2021b,Komossa2021c,Komossa2021d} have already been published until May 2022 \citep{Komossa2022a,Komossa2022b}. Data beyond May 2022 will be reported elsewhere. 
The Swift data is also consistent with the idea that the source of the flaring in the early part of 2022 was in the main jet \citep{Komossa2022a}.

It is not possible to say how the exact relation between the position of the secondary black hole in its orbit relative to the accretion disk is related to the level of activity seen in the optical light curve. There is of course an appropriate delay between a disk perturbation and its appearance in jet emission \citep{val17}. However, from the present work it appears that the relation was the same in the 2005 and 2022 impacts. 

\section*{Acknowledgements}
This work was partly funded by the NCN grant No. 2018/29/B/ST9/01793 (SZ) and JSPS KAKENHI grant No. 19K03930 (KM)

\section*{Data Availability}
The data published in this paper are available on reasonable request from the authors.
% \vspace{5mm}

% Don't change these lines
\bsp    % typesetting comment
\label{lastpage}
\end{document}